\documentclass[preprint, dvipdfmx]{ptephy_v1}
\usepackage{braket}
\usepackage{amsthm}
\usepackage{ascmac}
\usepackage{amssymb, amsmath}
\usepackage{bm}
\usepackage{physics}
\usepackage{graphicx}
\usepackage{docmute}
\usepackage{color}
\usepackage{autobreak}
\usepackage{url}
\usepackage{times}
\usepackage[T1]{fontenc}
\usepackage[normalem]{ulem}

\newcommand{\inner}[2]{\bm{#1}\cdot\bm{#2}} 
\newcommand{\mev}{{\mathrm{MeV}}}
\newcommand{\fm}{{\mathrm{fm}}}

\begin{document}
\title{Isotone Chain Study of $\bar{p}$-atom spectroscopy and \\ Strong Spin-orbit splittings}



\author{Kenta Yoshimura}

\author{Shunsuke Yasunaga}

\author{Daisuke Jido}

\author{Hiroyuki Fujioka}
\affil{Department of Physics, School of Science, Institute of Science Tokyo, Tokyo 152-8551, Japan}

\date{\today}

\begin{abstract}
Antiprotonic atoms have served as a pivotal tool for investigating the properties of baryon-baryon interactions, including their spin dependence.
Examining the spin-orbit splittings induced by their strong interactions also could help clarify the nature of the $\bar{p}$-nucleus interactions and their fraction mediated by scalar and vector mesons.
Although the strong spin-orbit splittings for a certain nucleus have been observed experimentally, thorough theoretical investigations have not yet been conducted.
In this study, theoretical calculations based on the Dirac equation are systematically performed for nuclei along several isotone ``chains''.
As a result, it is found that the magnitude of the strong spin-orbit splittings exhibits a significant dependence not only on the corresponding level shifts and widths almost linearly, but also on whether the optical potential enters as a vector or scalar potential.
A simple perturbative analysis indicates that the relativistic corrections have a dominant effect the magnitude of the splittings.
These results are expected to provide deeper insights into $\bar{p}$-nucleus interactions, and by extension baryon-baryon interactions, as well as into the properties of the mesons that mediate them.
\end{abstract}

\maketitle
\section{INTRODUCTION}\label{sec:intro}
Hadronic atoms refer to atomic systems in which mesons or baryons are bound to electronic orbits, earning more and more attention as an efficient tool for directly probing not only properties of the strong interactions, but also structures of atomic nuclei~\cite{batty1989a, wycech1996c, batty1997a, lubinski1998, schmidt2003,jastrzebski2004, klos2004,klos2007,wycech2007, friedman2007a,friedman2008,trzcinska2009, ficek2018,aumann2022}. 
For instance, recent successes of the piAF experiment, which measured the partial restoration of chiral symmetry by means of deeply bound pionic atoms~\cite{nishi2023}, have been remarkable, and further developments and applications are expected within various research realms.
Antiprotonic atoms are also a type of hadronic atoms~\cite{batty1989, klempt2005,  doser2022} and their spectroscopy is characterized by level shifts and widths stemming from the strong interaction~\cite{wycech1993,gotta1999, gotta2004}.
Regarding these spectral quantities, collaboration such as PS209 at LEAR have been conducted in the experimental domain~\cite{poth1985,trzcinska2001a, trzcinska2009}, while in the theoretical field considerable attempts have been made to provide comprehensive explanations for a wide range of nuclei by means of the ``global fitting'' of optical potential parameters~\cite{batty1981, cote1982, batty1987, batty1995b, friedman2004, friedman2005, friedman2013, friedman2014, friedman2015, friedman2019, friedman2019a}.

However, this endeavor involves significant difficulties; since these quantities strongly depend on the distance between the antiproton's orbit and the nucleus, the results can vary by orders of magnitude depending on the nucleus and observed orbit. 
In fact, it is already known that parameter sets obtained through global fitting, which are known as the Batty potential~\cite{batty1995b} or the Friedman potential~\cite{friedman2005}, do not reproduce experimental results in several nuclei~\cite{hartmann2001a, klos2007, yoshimura2024a}.
Additionally, it has been pointed out that the Friedman parameter underestimates the $\bar{p}$-nucleus annihilation cross section by a factor of 2 to 4 at low energies~\cite{friedman2014a}.
These facts clearly reflect that, to uncover the properties of antiprotonic atoms and strong interactions in more detail, it is crucial not merely to determine plausible values for the entire range of nuclei through statistical methods, but also to conduct detailed discussions on individual nuclides from various perspectives such as the parameter types, nuclear density profiles, and isotope or isotone dependence.
In our previous research, we have investigated the interrelation between the antiprotonic atom spectra of calcium isotopes and nuclear profiles~\cite{yoshimura2024a}.
In that work, we have proposed that, to explain the systematical differences in experimental results between $^{40}$Ca and $^{48}$Ca, the isovector term or p-wave term, although they have been previously considered non-dominant~\cite{friedman2014a}, be included with significant magnitude.
This finding provides unequivocal evidence that the $\bar{p}$-$A$ interaction exhibits isospin dependence, thereby strongly indicating that more detailed analysis specific to featured nuclei rather than mere global fitting is required.

In the present work, we will propose a novel observable to probe the $\bar{p}$-$A$ interaction, that is, the \textit{strong} spin-orbit splitting. 
By definition, this quantity represents the portion of the total spin-orbit splitting which originates from the strong interaction.
A key question is whether this contribution is sufficiently large to be experimentally accessible, where the electron orbit naturally undergoes splitting by spin-orbit coupling of the Coulomb force.
For instance, Ref.~\cite{kreissl1988a} reported that the strong spin-orbit splitting was observed in the antiprotonic spectrum of $^{174}$Yb.
However, because the deformation effects have been considered only approximately, the quantitative size of the effect remains uncertain.
On the other hand, Ref.~\cite{klos2004} argued that such strong spin-orbit effect was not observed in the spectra in $^{172}$Yb and $^{176}$Yb, raising the questions of why it would appear only in $^{174}$Yb among its isotopes, and what accounts for these differences.
To uncover the nature of the strong spin-orbit interactions, necessary is more wide-ranging and systematical investigations from both theoretical and experimental points of view.
Within the experimental fields, the applicability of the superconducting micro-calorimeter detector to $^{40}$Ca has been suggested in Ref.~\cite{higuchi2025}, where they have proposed that it provides a resolution of 50-70 eV in the energy range of interest, which should be sufficient to observe the spin-orbit splitting in eV-scale.
This fact indicates that near future experiments are expected to provide much more precise measurements of the antiprotonic spectra, where the spin-orbit splittings of many other nuclei would be clearly resolved.
This is why it would be strongly important to advance theoretical studies using more appropriate frameworks.
Under the present situation, many of the theoretical calculations of $\bar{p}$-$A$ spectra have incorporated the Klein-Gordon equation, in which the spin-orbit contribution is either neglected or treated only approximately.
In our previous research, we have verified that the spin $1/2$ Dirac equation can successfully describe the $\bar{p}$-Ca atom spectroscopy even when incorporating the effects of the anomalous magnetic moment and the optical potential~\cite{yoshimura2024a}.
The Dirac equation enables us to take into account the spin-orbit effects relativistically and should be a proper way to deal with the present issue.

To resolve the spin-orbit splitting experimentally, it is required that the splitting energy exceed the strong level width.
Given most of the splitting energy is attributed to the Coulomb force, having a higher proton number is advantageous in this respect.
On the other hand, as the proton number increases, the orbital radius contracts, enlarging the overlaps between wave functions of the antiproton and nucleus.
Because this effect broadens the level widths, higher proton numbers work against the observation in this respect.
For the target selection of nuclear species and orbits, one must carefully take this ``dilemma'' into consideration. 

This study focuses on the dependence of antiprotonic spectra on \textit{isotones} rather than isotopes.
In one isotone sequence, the neutron number remains fixed while only the proton number varies, thereby allowing the antiproton orbital radius to be more directly adjusted without much confounded by changes in nuclear size or the neutron skin.
By examining such isotone ``chains'', one can efficiently and accurately elucidate the systematical behaviour of spectral quantities across different nuclei and assess the feasibility of the strong spin-orbit splitting measurement.
In this paper, we mainly investigate the following two chains:
The $N=28$ isotones $^{48}$Ca, $^{50}$Ti, $^{52}$Cr, and $^{54}$Fe, the $N=50$ isotones $^{86}$Kr, $^{88}$Sr, $^{90}$Zr, and $^{92}$Mo.
Both these ``main'' chains have neutron magic number and thus it is expected that all these nuclei possess the spherical property, where we can assume the spherical symmetry.
Additionally in the latter sections we examine subsequent three chains; $N=32$ chain $^{58}$Fe, $^{59}$Co, and $^{60}$Ni, and $N=76$ isotones $^{114}$Cd, $^{115}$In, and $^{116}$Sn, and finally $N=126$ chain $^{208}$Pb and $^{209}$Bi.
These three ``sub''-chains contain odd nuclides and it is interesting to examine whether there is differences in the isotone dependence from the two main chains.


For the $\bar{p}$-nucleus interactions, the optical model is one common way to calculate the spectroscopic quantities.
Efforts to deal with the global fittings of the parametric coefficients to multiple nuclei have often been reported.
Prevailing one is Friedman's best fit $b_0 = 1.3+1.9i\,\fm$~\cite{friedman2005}, which contains only the isoscalar s-wave term depending on solely the nuclear density linearly.
Although it has been pointed out that this parameter does not reproduce experimental data for several specific nuclei, it is deemed adequate for investigating systematical trends in the spectra through sequences of nuclei in the present study.
There is a discourse on to what extent the optical potential enters as a vector potential or scalar potential~\cite{mishustin2005, lisboa2010, gaitanos2011}, which is dependent on which mesons mediate the interactions.
In this work, to simplify discussions, we employ only two cases; one is the case where all the optical potential enters as a vector potential, and the other is a scalar.
Using these models, we perform calculations for the aforementioned isotone chains to estimate the magnitude of the strong spin-orbit splittings, and uncover its systematicity.

This article is arranged as follows.
Section 2 provides the formalism of the used framework.
Section 3 demonstrates the calculation results and discusses them.
Section 4 summarizes the gist of this study and mentions the prospects.
The full calculation results in this work are shown in Appendix A.
The detailed analysis using a simple perturbative model is discussed in Appendix B.
\section{FRAMEWORK}
\subsection{Dirac Equation}
We consider the spherically symmetric $\bar{p}$-atom system. Each state is labeled by the total angular momentum $j$, the orbital angular momentum $l$, and the azimuthal angular momentum $m$. From the coupling of angular momentum with spin-$1/2$, two values of $l$ are possible for a given $j$. We use a shorthand $\pm$ to denote either sign: $l=j\pm1/2$ for fixed $j$, and $j=l\pm1/2$ for fixed $l$. Under the assumption of spherical symmetry, the Dirac spinor $\psi_{jm}^{\pm}(\bm{r})$ can be expressed as
\begin{equation}
    \psi_{jm}^{\pm}(\bm{r})=\mqty(i\frac{G_j^{\pm}(r)}{r}\mathcal{Y}_{jm}^{\pm}(\theta,\phi)\\\frac{F_j^{\pm}(r)}{r}(\inner{\sigma}{\hat{r}})\mathcal{Y}_{jm}^{\pm}(\theta,\phi)),
\end{equation}
where $G_j^{\pm}(r)$ and $F_j^{\pm}(r)$ are the radial wave functions of the large and small components, respectively, and $\mathcal{Y}_{jm}^{\pm}$ are the spherical harmonics with spin-1/2. The radial Dirac equation of the two-body systems at fixed energy $E$ can be written as
\begin{equation}
    \begin{aligned}
        (E - \mu- S(r) - V(r))G_j^{\pm}(r)
        &= -\dv{F_j^{\pm}(r)}{r}\pm\left(j+\frac{1}{2}\right)\frac{F_j^{\pm}(r)}{r} - \frac{\kappa}{2m}\dv{V}{r}F_j^{\pm}(r),\\
        (E + \mu + S(r) - V(r))F_j^{\pm}(r)
        &= \dv{G_j^{\pm}(r)}{r}\pm\left(j+\frac{1}{2}\right)\frac{G_j^{\pm}(r)}{r} - \frac{\kappa}{2m}\dv{V}{r}G_j^{\pm}(r).\label{Eq:RadialDirac}
    \end{aligned}
\end{equation}
Here, $S$ is the scalar potential, $V$ is the time component of the vector potential, $\mu$ is the reduced mass of the antiproton-nucleus system, and $\kappa$ is the anomalous magnetic moment of antiproton with $\kappa = 1.79284734$~\cite{smorra2017}.
The distinction between ``scalar'' and ``vector'' arises from whether the interaction responsible for generating the potential is mediated by a scalar meson or a vector meson.
Further details on notation are given in Ref.~\cite{yoshimura2024a}.
\subsection{Potentials}
The Coulomb potential consists of the pure Coulomb term and the vacuum polarization term:
\begin{equation}
    U_{\mathrm{Coul}}(r) = U_{\mathrm{PC}}(r) + U_{\mathrm{VP}}(r),
\end{equation}
where PC and VP stand for ``Pure Coulomb'' and ``Vacuum Polarization''.
The first term can be written as
\begin{equation}
    U_{\mathrm{PC}}(r) = -\alpha\int\dd^3\bm{r}' \frac{\rho_c(r')}{\abs{\bm{r}-\bm{r}^\prime}},
\end{equation}
with the finite size effect of the charge distribution.
The vacuum polarization is considered up to the second order in this work, where the detailed representation is given in Ref.~\cite{fullerton1976}.
With these formulae, the spin-orbit splitting for $n=11\to 10$ states of $\bar{p}$-$^{208}$Pb is calculated as $1199.88$ eV, which is greatly consistent with the experimental data $1199(5)$ eV~\cite{borie1983}.

The optical potential can be obtained via the linear density approximation, which results in
\begin{equation}
    \begin{aligned}
        U_\text{opt}(r) =&-\frac{4\pi}{2\mu}\qty(1+\frac{\mu}{M_N}) \qty[b_0\rho_0(r) + b_1\rho_1(r)]\\
        &+\frac{4\pi}{2\mu}\qty(1+\frac{\mu}{M_N})^{-1}\grad\qty[c_0\rho_0(r) + c_1\rho_1(r)]\cdot\grad,
    \end{aligned}\label{eq:opticalmodel}
\end{equation}
with the isoscalar and isovector densities written by linear combinations of neutron and proton densities $\rho_n$, $\rho_p$, as
\begin{eqnarray}
    \rho_0(r) &=& \rho_n(r) + \rho_p(r),\\
    \rho_1(r) &=& \rho_n(r) - \rho_p(r),
\end{eqnarray}
and with the averaged nucleon mass $M_N=938.918\,\mev$.
There have been many efforts to determine the paremetric coefficients $b_0$, $b_1$, $c_0$ and $c_1$.
Prevailing is the Friedman's best fit $b_0=1.3+1.9i\,\fm$~\cite{friedman2005}.
Within this fitting protocol, only the s-wave ($b_0$, $b_1$) terms have been taken into consideration and $b_1$ has been concluded to be negligibly small.
This is proportional to the isoscalar density and useful to investigate the systematical behaviour of the spectral quantities through isotone sequences.
In our previous research, we have suggested that the parameter sets incorporating the isovector or p-wave term added to Friedman's $b_0$ and verified its validity~\cite{yoshimura2024a}. 
In this study, however, these parameters are not employed, because they are deduced from the spectral properties of only Ca isotopes, as well as this study aims for extracting simple trends changing by various isotones.
Another choice is Wycech's best fit $b_0=1.8+0.95i\,\fm$ with $c_0=1.05i\,\fm$~\cite{wycech2007},
where the isovector terms ($b_1$, $c_1$ have been left out of the model.
The p-wave term depends on the derivative of the density and wave function, which is expected to exhibit more complex behaviour than the Friedman parameter through the sequences of nuclei.
This parameter will be also employed in the present study to confirm the trends of the calculated results are common.

To examine the systematics of the strong spin-orbit splittings, it is better to check both the two cases where the optical potential enters as the vector and scalar potentials.
Although the difference is the change of sign in the contribution to the large $F_{lj}$ component in Eq.~\eqref{Eq:RadialDirac}, as is mentioned later this difference significantly alters the magnitude of the strong spin-orbit splittings as their relativistic corrections.
One standard choice is to set all the optical potential as a scalar potential, that is $S(r) = U_\text{opt}(r)$ and $V
(r) = U_\text{Coul}(r)$.
However, in this work, to explore the roles played by the scalar and vector optical potential, we also examine the case where all the optical potential enters as a vector potential, thus $S(r) = 0$ and $V(r) = U_\text{Coul}(r) + U_\text{opt}(r)$.
It is expected that the actual $\bar{p}$-nucleus interaction is a mixture of scalar and vector components; however, in the present study, we restrict ourselves to the above two cases in order to examine the role played by each component individually.

\subsection{Densities}
We need to care the treatment of the nuclear densities.
This is because, while the proton density distributions have been accurately measured via the electron scattering or muonic atom experiments~\cite{devries1987, fricke1995}, there is poor information on the density of neutrons.
This is why simple phenomenological models are often employed such as the 2-parameter Fermi model:
\begin{equation}
    \rho_q(r) = \frac{\rho_c}{1 + \exp((r-R_q)/a_q)},
\end{equation}
with parametric coefficients $R_q$ and $a_q$.
In this study, the parameter $R_p$ is employed from the muonic experimental data with $a_{q}$ fixed to $a_n=a_p = 2.3\,\fm/(4\ln 3)$ fm~\cite{fricke1995}.
The neutron radius parameter $R_n$ is determined to reproduce the neutron skin thickness $\Delta r_{np}$ deduced from the antiprotonic atom experiment~\cite{jastrzebski2004}, which is written as
\begin{equation}
    \Delta r_{np} = -0.09 + 1.46\delta,
\end{equation}
as a function of $\delta = (N_n - N_p) / (N_n + N_p)$.
To take into account the finite size effect of nucleons, it is necessary to incorporate the density folding with the bare particle density~\cite{friedman2005}.
The 2-parameter Fermi model is, however, constructed for describing the nuclei's charge density instead of particle density, so that we can use these parameters as themselves.

In Table \ref{tab:parameters} shown are the nuclear mass and profile parameter sets for the nucleus belonging to the two main chains proposed in Sec. \ref{sec:intro}.
The masses of nuclei are summarized in Ref.~\cite{zotero-626}, and $R_p$ is taken from Ref.~\cite{fricke1995}.
In the table there are also calculated proton and neutron mean-square radii $\langle r_q \rangle$ and their differences $\Delta r_{np} =\langle r_n \rangle - \langle r_p \rangle$.

\begin{table}[tbp]
    \centering
    \caption{The parameters used for nuclear mass $M_{\mathrm{Nucl}}$, proton radius $R_p$ and neutron radius $R_n$ alongside the calculated values of the proton root-mean-square radius $\langle r_p\rangle$, neutron radius $\langle r_n \rangle$ and their difference $\Delta r_{np}$.}
    \begin{tabular*}{0.9\columnwidth}{@{\extracolsep{\fill}}c|cccccc}
       \hline Nucl & $M_{\mathrm{Nucl}}\,(\mev)$ & $R_p\,(\fm)$ & $R_n\,(\fm)$ & $\langle r_p\rangle\,(\fm)$ & $\langle r_n \rangle\,(\fm)$ & $\Delta r_{np}\,(\fm)$ \\
        \hline\hline
        $N=28$\\
        $^{48}$Ca & 44657.3 & 3.7231 & 3.9596 & 3.479 & 3.632 & 0.153\\
        $^{50}$Ti & 46522.5 & 3.8659 & 3.9964 & 3.571 & 3.656 & 0.085\\
        $^{52}$Cr & 48381.4 & 3.9742 & 4.0083 & 3.642 & 3.664 & 0.022\\
        $^{54}$Fe & 50243.6 & 4.0546 & 4.0546 & 3.694 & 3.694 & 0.0\\
        \hline 
        $N=50$\\
        $^{86}$Kr & 80023.8 & 4.7819 & 4.9959 & 4.184 & 4.331 & 0.147\\
        $^{88}$Sr & 81882.1 & 4.8399 & 4.9980 & 4.224 & 4.333 & 0.109\\
        $^{90}$Zr & 83744.2 & 4.9011 & 5.0057 & 4.266 & 4.338 & 0.072 \\
        $^{92}$Mo & 85609.2 & 4.9754 & 5.0288 & 4.317 & 4.354 & 0.037\\
        \hline\hline
    \end{tabular*}
    \label{tab:parameters}
\end{table}

Incorporating the Coulomb potential and the optical potential by means of the density models, we have the Dirac Hamiltonian operating the wave functions $F(r)$ and $G(r)$.
Diagonalizing the Hamiltonian matrix with fixed (large component's) orbital angular momentum $l$ and total angular momentum $j$, we have a set of energy eigenvalues $E_{lj}$ and wave functions $F_{lj}(r)$ and $G_{lj}(r)$.
The binding energies can be calculated by subtracting the mass term from the obtained eigenvalues, written as
\begin{equation}
    B_{lj} = -(E_{lj}- \mu) .
\end{equation}
The binding energy may be decomposed into the (i) bare binding energy contributed from the Coulomb potential, (ii) strong shift $\varepsilon$ and (iii) level width $\Gamma$, explicitly written down as
\begin{equation}
    B_{lj} = B_{lj}^{(\text{em})} - \varepsilon_{lj} - \frac{\Gamma_{lj}}{2}i.
\end{equation}
The electromagnetic and strong contributions to the real part are separated out using the binding energies in the case without the optical potential $E^{(\text{em})}_{lj}$, according to
\begin{equation}
    B_{lj}^{(\text{em})} = - (E^{(\text{em})}_{lj} - \mu).
\end{equation}

We additionally mention the spin-orbit splittings.
Theoretically the spin-orbit splittings can be obtained as the difference of binding energies between spin states;
\begin{equation}
    \Delta B_{l} = B_{l-} - B_{l+},
\end{equation}
where we remind that the suffices $\pm$ stand for the total angular momenta with spin states $j = l \pm 1/2$ for fixed $l$.
In experiments, observable are the transition energies between two orbits, namely the practical spin-orbit splitting is 
\begin{equation}
    \Delta B_{l+1\to l} = \Delta B_{l} - \Delta B_{l+1}.
\end{equation}
This obtained spin-orbit splittings can be further decomposed into the contribution from the Coulomb potential 
\begin{equation}
    \Delta B^{(\text{em})}_{l+1\to l} = \Delta B^{(\text{em})}_{l} - \Delta B^{(\text{em})}_{l+1}
\end{equation}
and from the strong interactions which can be written as
\begin{eqnarray}
    \Delta \varepsilon_{l+1\to l} = \Delta\varepsilon_{l} -\Delta \varepsilon_{l+1},
\end{eqnarray}
with
\begin{eqnarray}
    \Delta B^{(\mathrm{em})}_{l} &=& B^{(\mathrm{em})}_{l-} - B^{(\mathrm{em})}_{l+},\\
    \Delta\varepsilon_{l} &=& \varepsilon_{l-}-\varepsilon_{l+}.
\end{eqnarray}
These values in general should vary depending not only on the potential shape but also on whether the optical potential enters as the scalar and vector potential.
By performing calculations for prepared isotone chains, it should be possible to extract the systematical behaviour of these spectral quantities.
\section{RESULTS}

\subsection{Shifts and Widths}

Table 2 lists the strong shifts, level widths, and magnitudes of the spin-orbit splitting for all examined nuclei in the main-chains.
Both the shifts and widths are presented for the case $j=l+1/2$. 
In each instance, we confirm that the results for $j=l-1/2$ do not differ significantly.
The full result including the $j=l-1/2$ case is displayed in Appendix A.
Furthermore, for those nuclei which have been already measured experimentally, the corresponding experimental values are shown to the right of the calculated values~\cite{trzcinska2001a}. 
For the spin-orbit splitting, three sets of results are provided: one for the case involving only the Coulomb force $\Delta B^{(\mathrm{em})}_{l+1\to l}$ and another for the case in which the optical potential contributes, subdivided into scenarios where the optical potential enters as the vector $\Delta \varepsilon^V_{l+1\to l}$ or as the scalar potential $\Delta \varepsilon^S_{l+1\to l}$.
We here note that, as is mentioned before, actual $\bar{p}$-$N$ interactions contains both scalar and vector components in some proportion and these results are limited in only the simplified cases.

\begin{table}[tbp]
    \centering
    \caption{The calculation results of the binding energy $B^{\text{(em)}}_{l+}$, strong shift $\varepsilon_{l+}$, level width $\Gamma_{l+}$, spin-orbit splittings attributed to the Coulomb potential $\Delta B^{(\mathrm{em})}_{l+1\to l}$, optical potential in the case with the vector potential $\Delta \varepsilon^V_{l+1\to l}$ and the scalar case $\Delta \varepsilon^S_{l+1\to l}$. All physical quantities are written in a unit of eV, except for the binding energy in keV. The question mark attached to the experimental width of $^{48}$Ca indicates that the value was obtained as an average across its isotopes $^{40,42,43,44,48}$Ca, rather than being specific to $^{48}$Ca}
    \begin{tabular*}{\columnwidth}{@{\extracolsep{\fill}}c|c|cccccccc}
        \hline\hline 
         &&$B^{\text{(em)}}_{l+}$& \multicolumn{2}{c}{$\varepsilon_{l+}$ (eV)} & \multicolumn{2}{c}{$\Gamma_{l+}$ (eV)} & &&\\
        \cline{4-5}\cline{6-7}Nucl & $(n,l)$&  (keV)& calc & exp & calc & exp & $\Delta B^{\text{(em)}}_{l+1\to l}$ & $\Delta \varepsilon^V_{l+1\to l}$ & $\Delta \varepsilon^S_{l+1\to l}$\\ 
        \hline
        $N=28$&$(5,4)$&&&&\\
        $^{48}$Ca && $392.9$ & $0.99$ & $33(12)$ & $39.67$ & $35(6)?$ & $238.78$ & $-0.313$ & $0.042$ \\
        $^{50}$Ti && $475.9$ & $5.091$ &  & $103.9$ &  & $350.21$ & $-1.009$ & $0.015$ \\
        $^{52}$Cr && $566.9$ & $17.49$ &  & $242.4$ &  & $496.81$ & $-2.816$ & $-0.233$ \\
        $^{54}$Fe && $665.9$ & $50.84$ & $155(60)$ & $523.7$ & $545(45)$ & $685.30$ & $-7.201$ & $-1.228$ \\
        \hline $N=50$&$(6,5)$&&&&\\
        $^{86}$Kr && $892.7$ & $22.77$ &  & $308.3$ &  & $877.09$ & $-4.072$ & $-0.815$ \\
        $^{88}$Sr && $995.0$ & $49.21$ &  & $540.2$ &  & $1089.9$ & $-8.136$ & $-2.076$ \\
        $^{90}$Zr && $1102.9$ & $100.1$ & $28(29)$ & $915.0$ & $1040(40)$ & $1339.3$ & $-15.60$ & $-4.769$ \\
        $^{92}$Mo && $1216.4$ & $196.6$ &  & $1513$ &  & $1629.2$ & $-29.09$ & $-10.27$ \\
        \hline\hline
    \end{tabular*}
    \label{tab:shift_and_width}
\end{table}

From the comparison between the calculated and experimental results, two contrast characteristics are made clear. 
On the one hand, with respect to the level width, the calculations generally reproduce the experimental values for almost all listed nuclei. 
This finding is consistent with a previous research indicating that the fitting protocol for the periphery of Ca isotopes based on the Friedman potential has yielded outcomes agreeing with experimental data~\cite{hartmann2001a}. 
On the other hand, for the level shift, none of the nuclei with known experimental values are reproduced correctly. 
In particular, it is evident that the results for $^{48}\mathrm{Ca}$ and $^{90}\mathrm{Zr}$ deviate from the experimental values by more than $2\sigma$. 
This discrepancy underscores the necessity of not only examining global behaviour but also conducting more detailed investigations focused on individual nuclei (for an attempt involving Ca isotopes, see Ref.~\cite{yoshimura2024a}).

Turning to the strong spin-orbit splitting, two main tendencies can be discerned. 
In the first place, the magnitude of the strong spin-orbit splitting is clearly larger when the strong interaction is implemented as a vector potential, rather than as a scalar potential. 
This behaviour can be easily explained from the non-relativistic expansion of the Dirac equation.
In the expansion of $1/m$, the Dirac Hamiltonian can be written as
\begin{equation}
    H^\prime = \beta(m+S) + V - \frac{1}{4m^2}\frac{1}{r}\frac{\dd}{\dd r} ( \beta S - V) \bm{L}\cdot\bm{S} - \frac{1}{8m^2}[\grad^2 (\beta S + V)] + \cdots,\label{eq:nonrelaDirac}
\end{equation}
where the third term corresponds to the spin-orbit interactions.
Because $\beta$ provides a factor $1$ for the large component $G(r)$, the spin-orbit term clearly depends on the $(S-V)$ and in this case both Coulomb and optical potentials are always attractive.
This is why the optical potential entering as the scalar potential cancels with the contribution from the Coulomb potential each other. 
In the second place, in both cases of vector and scalar potential, the magnitude of strong spin-orbit splittings seems to scale up as the corresponding strong shifts and level widths.
In the following sections, let us delve deeper into the correlations among the spectral quantities.

\subsection{Level Widths Dependencies}
\begin{figure}[tbp]
    \centering
    \includegraphics[width=\columnwidth]{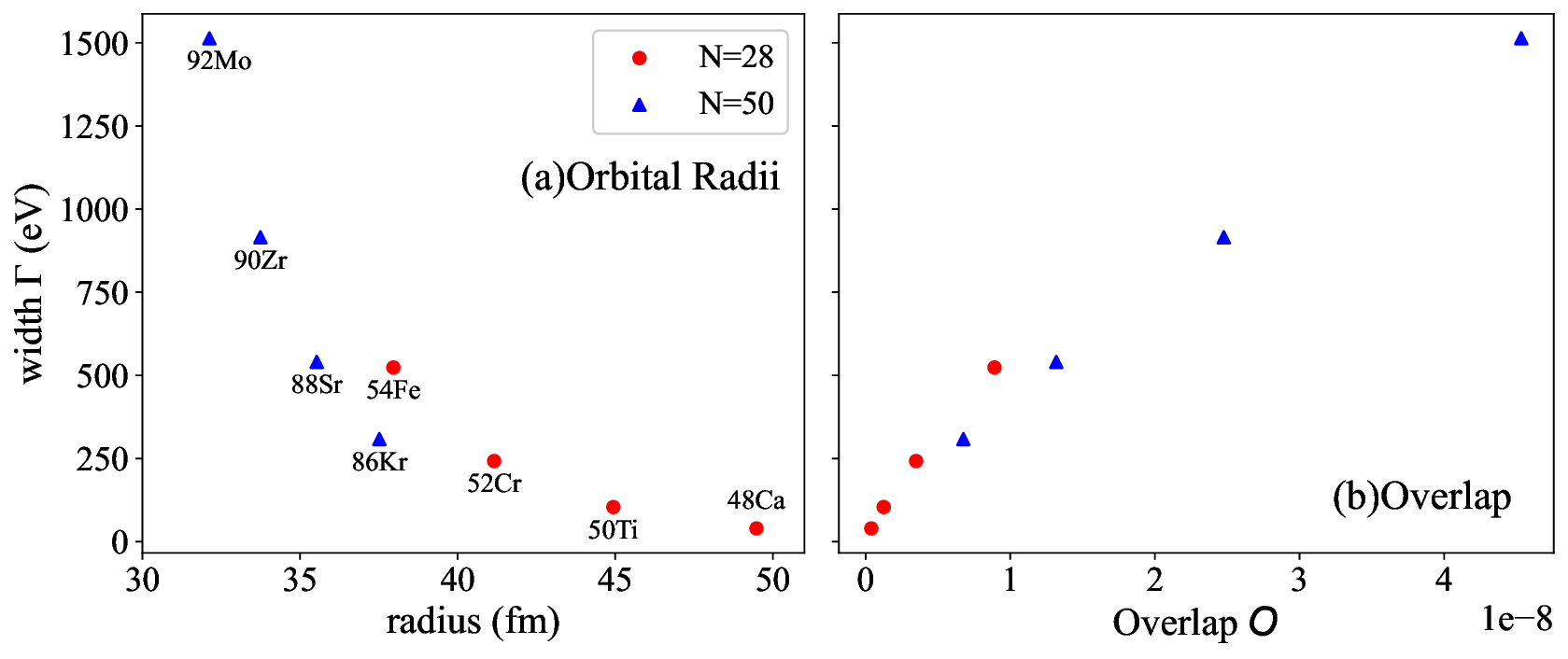}
    \caption{The level widths of the main-chain nuclei as functions of their (a) charge radii and (b) wave function overlaps with the optical potential. In the left panel, the target nucleus is specified for each plotted point.}
    \label{fig:overlap_chargeradius}
\end{figure}

Figure \ref{fig:overlap_chargeradius} examines the correlation between the level width and the species of antiprotonic atom, taking into consideration only the nuclei belonging to the main-chains.
In the left panel the level width is plotted as a function of the orbital radius of the antiproton, whereas in the right panel it is shown as a function of the overlap between the wave-function and the nuclear density;
\begin{equation}
    \mathcal{O} = \int\dd r \,4\pi r^2 \rho(r)\qty[\abs{F(r)}^2 + \abs{G(r)}^2].
\end{equation}
As can be seen from this figure, there is a clear correlation between the level width, the orbital radius, and the overlap with the wave function. That is, although the behavior of the level width varies somewhat depending on the orbits, it is generally found to be inversely proportional to the orbital radius and linearly proportional to the overlap.
The latter point has also been mentioned in the previous work where they evaluate spectral quantities with simple model calculations~\cite{gustafsson2025}.
This suggests that spectral quantities such as the level width implicitly include effects related to the correlation between the wave function and its density.
Therefore, for strong spin-orbit splitting as well, since it arises due to the influence of the optical potential, one should expect to find similar correlations with quantities such as the shift and width, which will be thoroughly examined later.

\subsection{Strong Splittings}

\begin{figure}[tbp]
    \centering
    \includegraphics[width=\columnwidth]{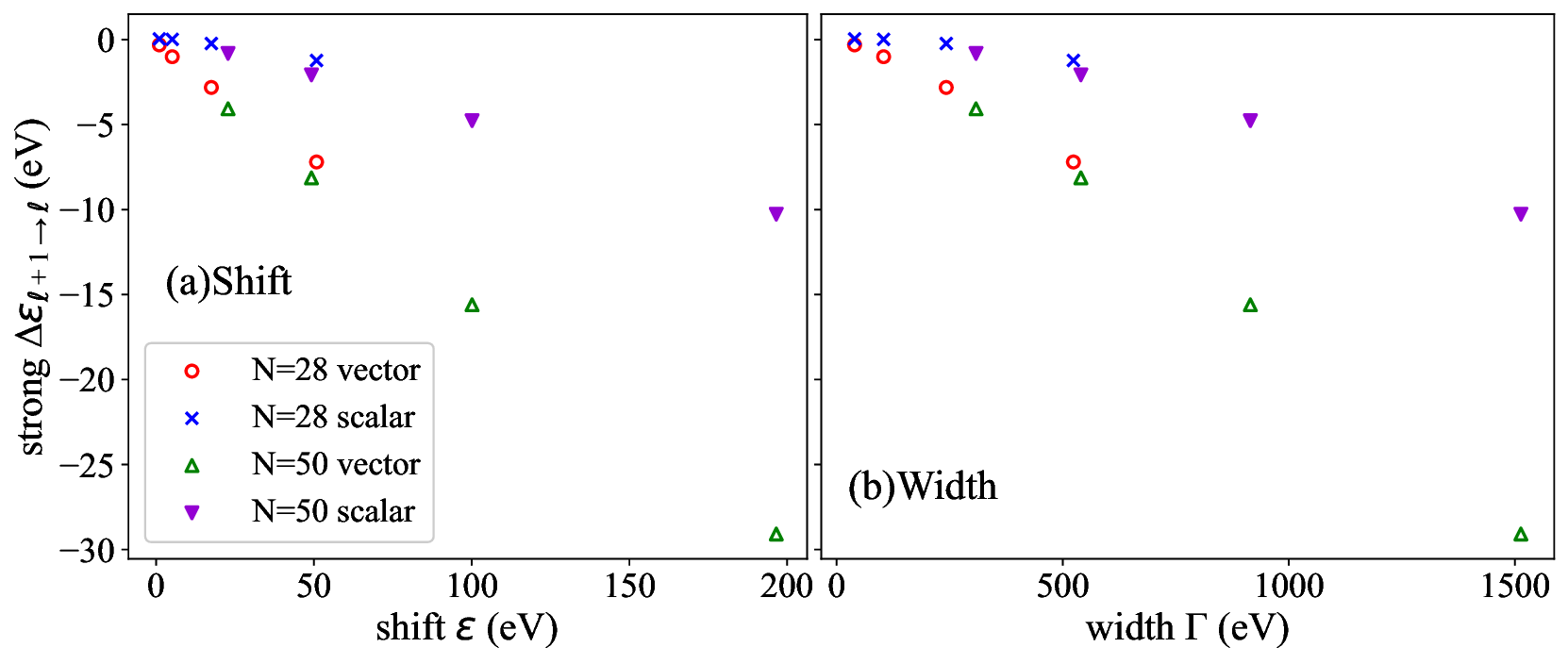}
    \caption{The calculated strong spin-orbit splitting as a function of the level shifts (left panel) and widths (right panel). In both panels, results for $N=28$ chain nuclei with vector potential, $N=28$ with scalar potential, $N=50$ with vector potential, $N=50$ with scalar potential are indicated by red circles, blue crosses, green upward triangles, violet downward triangles, respectively.}
    \label{fig:shiftandwidth}
\end{figure}

Figure \ref{fig:shiftandwidth} plots the strong spin-orbit splitting as a function of the shift (left panel) and the width (right panel), for the main-chain nuclei.
In both figures, used symbols are as follows: red circles for the $N=28$ chain nuclei with a vector potential, blue crosses for a scalar potential, green upward triangles for the $N=50$ chain with a scalar potential and purple downward triangles for the scalar.
These plots demonstrate that the strong spin-orbit splitting shows a profound linear correlation with both the level widths and strong shifts.
Although this is a natural consequence in the sense that all these quantities are governed by the overlaps between the wave functions and optical potential, it is intriguing that its dependence is so strong as to be linear.

\begin{figure}[tbp]
    \centering
    \includegraphics[width=\columnwidth]{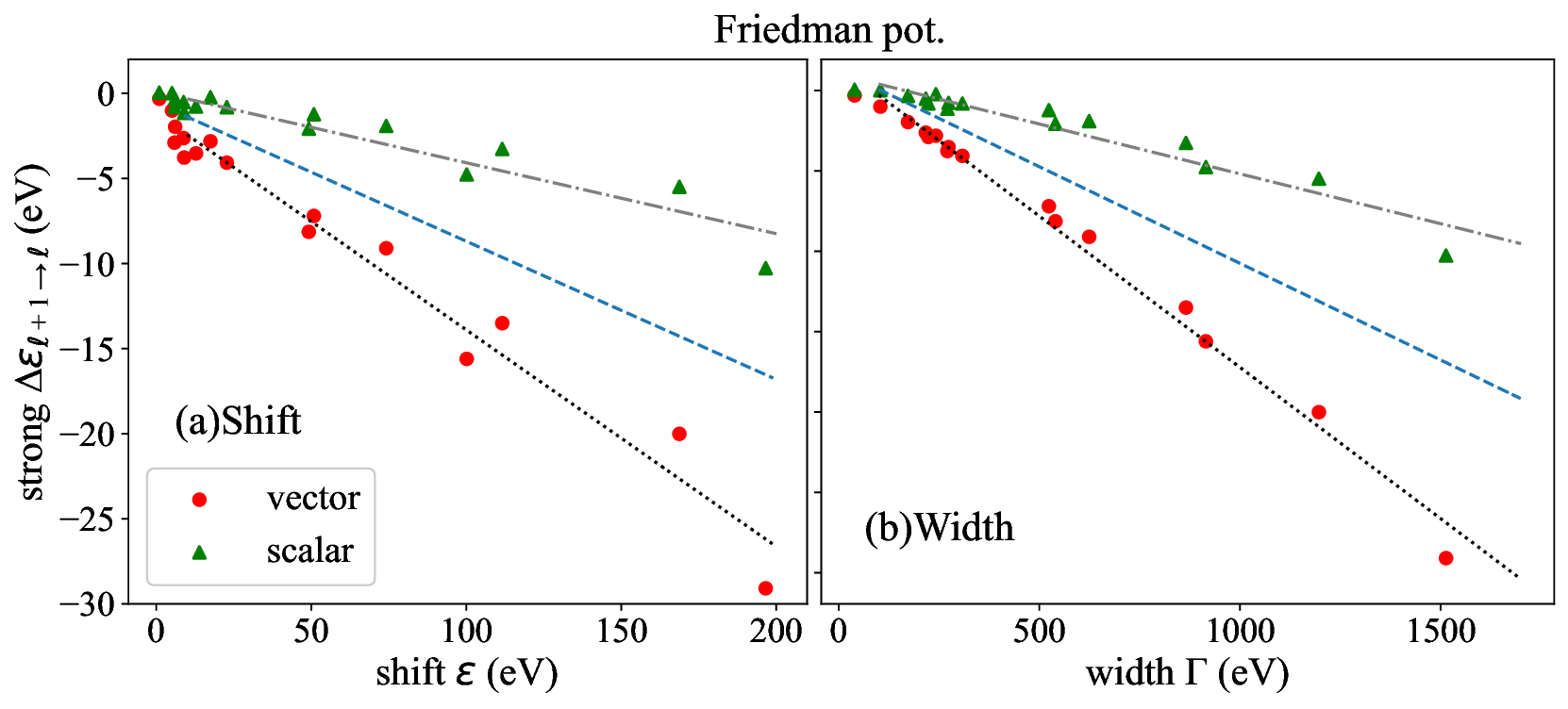}
    \caption{Calculated strong spin-orbit splitting as a function of the (a) strong shift and (b) level width. In both panels, results in the case where the optical potential enters as a vector (scalar) potential are indicated by red circles (green triangles). For the optical potential the Friedman potential is used. The black dotted and dash-dotted lines indicate the results of the linear fitting for the vector and scalar sequences, while the blue dashed line shows the prediction when the scalar-vector proportion in the optical potential is in accord with the standard relativistic mean-field (RMF) model.}
    \label{fig:shift_width_split_swave}
\end{figure}

\begin{figure}[tbp]
    \centering
    \includegraphics[width=\columnwidth]{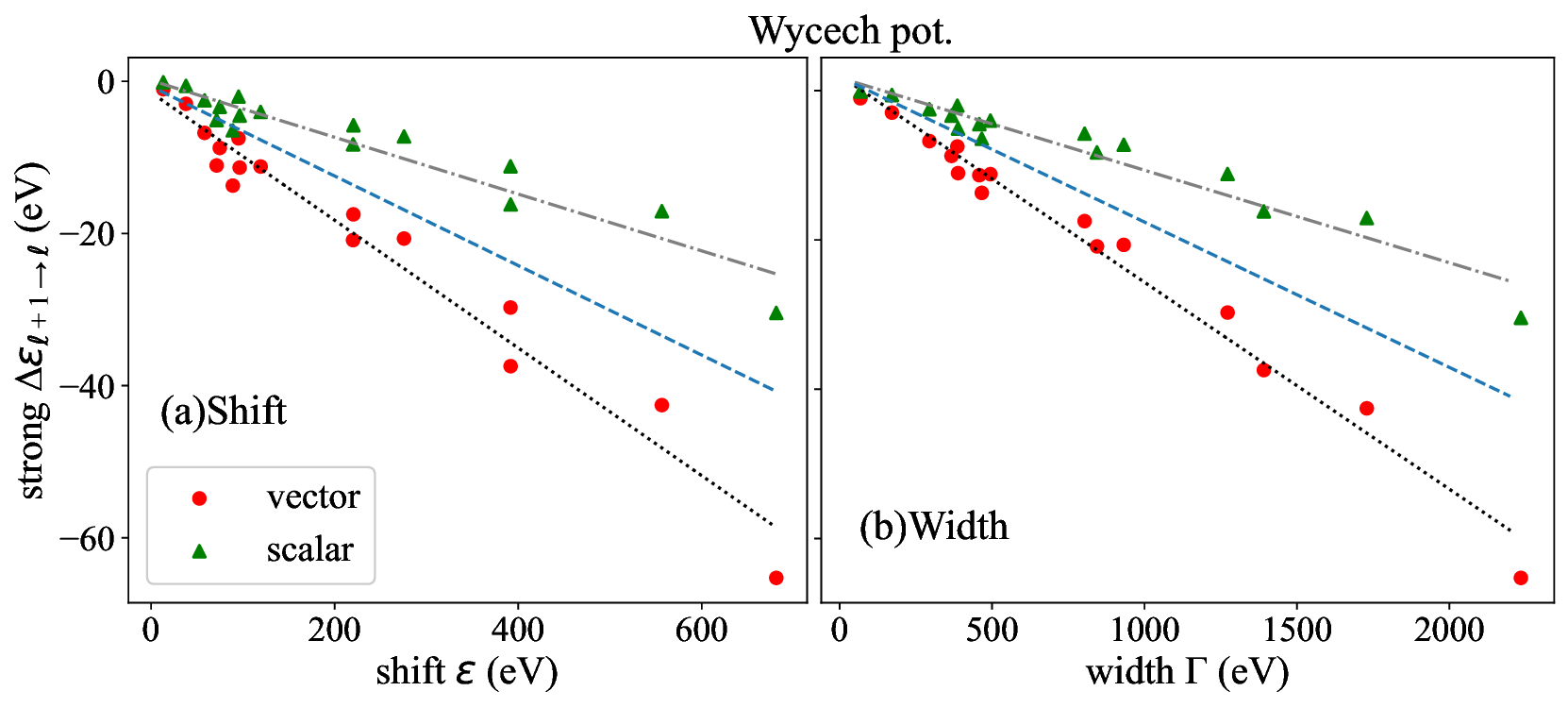}
    \caption{Same with Fig.~\ref{fig:shift_width_split_swave}, but for the Wycech potential.}
    \label{fig:shift_width_split_pwave}
\end{figure}

To verify this conclusion and further enhance the statistical reliability, we examine several additional calculations to be performed, including the nuclei belonging to the sub-chains.
Additionally, alongside the commonly used Friedman parameters, the Wycech parameters incorporating a p-wave term are also employed. 
The p-wave term, involving derivatives of the wave function and nuclear density, is introduced to better capture surface effects of the nucleus and can thereby lead to more complex behaviour of the spectra. 
In Figure \ref{fig:shift_width_split_swave} and \ref{fig:shift_width_split_pwave}, the same plots with Fig.~\ref{fig:shiftandwidth} but including the sub-chain nuclei for s-wave and p-wave parameters, respectively.
In both panels, red circles are results for the case where the optical potential enters as a vector potential, while green triangles indicate the scalar potential results.
From these results it is found that, notwithstanding the introduction of new chains and the inclusion of a p-wave term, a clear and nearly linear correlation with the strong spin-orbit splittings emerges regarding to both the strong shifts and level widths.
As is discussed in Appendix B, this behaviour is supported from the perturbative analysis using the simple non-relativistic model.


\begin{table}[h]
    \centering
    \caption{The results of $\chi^2$ fitting with a linear function $y=ax+b$ with parametric coefficients $a$ and $b$, the level shifts or widths $x$ and the strong spin-orbit splittings $y$.
    The results in cases with vector and scalar potentials are indicates in the left and right side of the table, respectively.}
        \begin{tabular}{c c c | c c c}
            \hline
            vector & $a$ & $b$ &
            scalar & $a$ & $b$ \\
            \hline\hline
            Friedman pot.&&&&&\\
            shift & $-0.127$ & $-1.17$  &
            shift & $-0.042$ & $0.090$  \\
            width & $-0.019$ & $1.59$ &
            width & $-0.006$ & $1.01$ \\
            \hline 
            Wycech pot.&&&&&\\
            shift & $-0.084$ & $-1.47$  &
            shift & $-0.037$ & $0.114$  \\
            width & $-0.028$ & $2.03$  &
            width & $-0.012$ & $1.72$  \\
            \hline
        \end{tabular}
    \label{tab:fitresult}
\end{table}

In both figures, the black dotted and dot-dashed lines attached along the vector and scalar sequences indicate the results of the linear fitting to them,
while Table \ref{tab:fitresult} shows the result of the coefficient values.
The fitting function is $y=ax+b$ with parameters $a$ and $b$, the level shifts (widths) $x$, and the strong spin-orbit splittings $y$.
The table reveals that in every instance a linear regression furnishes an good representation of the data. 
Moreover, for all cases examined, the fit to the level width is consistently superior to that obtained for the strong shift. 
These findings indicate that the level width is the more suitable predictor for the magnitude of the strong spin-orbit splitting. 

From these results, two conclusions follows.
In the first place, there is a definite linear correlation between the level width and the magnitude of the strong spin-orbit splittings, which means that we can anticipate the sequence of the splittings through multiple experiments or straightforward model calculations.
In the second place, their dependence significantly differs according to whether the optical potential is introduced as a vector or scalar potential.
Consequently, given the measurement on the sequence of strong spin-orbit splittings, one can infer the respective proportions of the vector and scalar components in the $\bar{p}$-nucleus optical potential, that is, the relative roles played by scalar and vector mesons in mediating the $\bar{p}$-nucleus interactions.
As an example, let us consider the case of the standard relativistic mean field (RMF) model at saturation density, where the scalar and vector potentials take values of $S=-350\,\mev$ and $V=300\,\mev$, respectively~\cite{COHEN1995221}.
For antiprotons, the sign of the vector potential is reversed, resulting in $V=-300\,\mev$~\cite{friedman2005,HRTANKOVA2016197}.
Figure 3 and 4 show the dependence of the strong spin-orbit splittings in the case of such a potential ratio as blue dashed lines.
It can be seen from that the result lies approximately midway between the two fitted lines.
This indicates that, if we assume that a linear dependence of the strong spin-orbit splitting on the level shifts and widths is robust, then there is a one-to-one correspondence between scalar-vector composition and the observed splitting behaviour.
In this case, it becomes possible to infer the scalar-vector composition from a sequence of a few experimental measurements.
In reality, it is also conceivable that the strong spin-orbit splitting may deviate from such a linear dependence.
In such cases, it would imply the presence of novel properties that are not captured by the conventional global fitting approach.
In this sense as well, the observation of a linear relationship between the strong spin-orbit splittings and widths or shifts is of particular significance.
These conclusions indicate that the strong spin-orbit splittings would be served as a novel tool for providing new perspectives on the $\bar{p}$-nucleus interactions and consequently strong interactions.

\section{SUMMARY}
In this study, we have carried out systematic calculations of the strong spin-orbit splittings in antiproton atomic spectra, based on the spin-dependent relativistic Dirac equation. 
After pointing out the ``dilemma'' in the observability of the spin-orbit splittings stemming from a trade-off with the level width, we have proposed a novel approach for estimating the appropriate orbital radius by systematically performing calculations along isotone chains.
As a simplified model for the optical potential, we examined two limiting cases: one in which the entire potential enters as a scalar field and another in which it is treated as a vector field. 

The analysis yielded several findings with important implications for future investigations of antiproton atoms.
In the first place, we have found a pronounced linear correlation between the magnitude of the strong spin-orbit splitting and both the level shifts and the widths. 
Consequently, this correlation enables the prediction of the strong spin-orbit magnitudes directly from measured widths without performing a large-scale global calculation or experiment.
In the second place, the magnitude of the strong spin-orbit splitting differs significantly depending on whether the optical potential is introduced as a scalar or as a vector field. 
If experimentally measured, this sensitivity would help determine the fraction of the $\bar{p}$-$A$ interaction mediated by scalar or vector mesons. 
In addition to these findings, it has been made clear that, for several nuclei examined, the calculated strong shifts deviate from available experimental data.
This result suggests that a correct description of antiproton spectra requires investigations which go beyond the conventional global fitting procedure and incorporate other perspectives.

These results offer deeper insight into antiproton atoms and $\bar{p}$-$A$ interactions. 
The experimental observation of strong spin-orbit splitting would mark a major milestone in this field. 
Improving the theoretical precision will require more sophisticated models for level shifts and widths.
Moreover, identifying the scalar and vector components of the $\bar{p}$-$A$ interaction will be crucial for constructing self-consistent potentials, such as those employed in relativistic mean-field (RMF) theories.
By approaching strong spin-orbit splitting from this new perspective, we may gain a clearer understanding of antiproton interactions—and, more broadly, baryon-baryon interactions in general.

\section*{ACKNOWLEDGEMENT}
One of the author K.Y. acknowledge the financial support from the JSPS Research Fellow, Grant No. JP24KJ1110.
The work of S.Y. was partly supported by JST SPRING, Grant Number JPMJSP2180.
The work of D.J. was partly supported by Grants-in-Aid for Scientific Research from JSPS (JP22H04917, JP23K03427, JP25K07315).
This work mainly made use of computational resources of the Yukawa-21 supercomputer at Yukawa Institute for Theoretical Physics (YITP), Kyoto University.

\appendix 

\section{Full Calculation Results}
Tables \ref{tab:fullresultfriedman} and \ref{tab:fullresultwycech} show the full calculation results of the strong shifts, level widths, and magnitudes of electromagnetic and strong spin-orbit splittings in unit of eV.
Table \ref{tab:fullresultfriedman} shows the results in the case when the Friedman parameter is used, while Table \ref{tab:fullresultwycech} corresponds to the Wycech parameter case.
Both tables include the calculation results for the sub-chain nuclei.

\begin{table}[tbp]
    \centering
    \caption{Full calculation results of the strong shifts $\varepsilon_{l+}$, $\varepsilon_{l-}$, level widths $\Gamma_{l+}$, $\Gamma_{l-}$, electromagnetic and strong spin-orbit splittings $\Delta B$, $\Delta \varepsilon_V$, and $\Delta \varepsilon_S$. For the optical potential, the Friedman parameter is used.}
    \begin{tabular*}{\columnwidth}{@{\extracolsep{\fill}}c|c|ccccccc}
        \hline\hline
        Nucl & $(n,l)$ & $\varepsilon_{l+}$ & $\varepsilon_{l-}$ & $\Gamma_{l+}$ & $\Gamma_{l-}$ & $\Delta B^\text{(em)}_{l+1\to l}$ & $\Delta \varepsilon^V_{l+1\to l}$ & $\Delta \varepsilon^S_{l+1\to l}$ \\
        \hline
        $^{48}\mathrm{Ca}$ & $(5,4)$ & $0.99$ & $1.30$ & $39.67$ & $41.87$ & $238.78$ & $-0.313$ & $0.042$ \\
        $^{50}\mathrm{Ti}$ &  & $5.09$ & $6.10$ & $103.91$ & $109.87$ & $350.21$ & $-1.009$ & $0.015$ \\
        $^{52}\mathrm{Cr}$ &  & $17.49$ & $20.3$ & $242.36$ & $256.72$ & $496.81$ & $-2.816$ & $-0.233$ \\
        $^{54}\mathrm{Fe}$ &  & $50.84$ & $58.04$ & $523.74$ & $555.54$ & $685.30$ & $-7.201$ & $-1.228$ \\
        \hline $^{86}\mathrm{Kr}$ & $(6,5)$ & $22.77$ & $26.84$ & $308.30$ & $330.58$ & $877.093$ & $-4.072$ & $-0.815$ \\
        $^{88}\mathrm{Sr}$ &  & $49.21$ & $57.35$ & $540.19$ & $580.07$ & $1089.89$ & $-8.136$ & $-2.076$ \\
        $^{90}\mathrm{Zr}$ &  & $100.14$ & $115.76$ & $915.00$ & $983.70$ & $1339.30$ & $-15.60$ & $-4.769$ \\
        $^{92}\mathrm{Mo}$ &  & $196.58$ & $225.72$ & $1513.4$ & $1628.2$ & $1629.2$ & $-29.09$ & $-10.271$ \\
        \hline $^{58}\mathrm{Fe}$ & $(5,4)$ & $74.18$ & $83.29$ & $624.23$ & $660.44$ & $686.14$ & $-9.103$ & $-1.912$ \\
        $^{59}\mathrm{Co}$ &  & $111.61$ & $125.12$ & $865.39$ & $916.41$ & $798.44$ & $-13.51$ & $-3.267$ \\
        $^{60}\mathrm{Ni}$ &  & $168.74$ & $188.78$ & $1196.6$ & $1267.9$ & $923.99$ & $-20.01$ & $-5.493$ \\
        \hline $^{114}\mathrm{Cd}$ & $(7,6)$ & $6.092$ & $8.051$ & $172.75$ & $187.96$ & $1131.8$ & $-1.965$ & $-0.336$ \\
        $^{115}\mathrm{In}$ &  & $8.828$ & $11.45$ & $217.21$ & $236.55$ & $1229.55$ & $-2.626$ & $-0.512$ \\
        $^{116}\mathrm{Sn}$ &  & $12.82$ & $16.34$ & $273.74$ & $298.36$ & $1333.55$ & $-3.526$ & $-0.775$ \\
        \hline $^{208}\mathrm{Pb}$ & $(9,8)$ & $5.891$ & $8.767$ & $223.48$ & $250.03$ & $2232.1$ & $-2.895$ & $-0.807$ \\
        $^{209}\mathrm{Bi}$ &  & $9.007$ & $12.76$ & $271.01$ & $303.24$ & $2343.7$ & $-3.771$ & $-1.142$ \\
        \hline\hline
    \end{tabular*}
    \label{tab:fullresultfriedman}
\end{table}

\begin{table}[tbp]
    \centering
    \caption{The same with Table \ref{tab:fullresultfriedman}, but for the Wycech potential.}
    \begin{tabular*}{\textwidth}{@{\extracolsep{\fill}}c|c|ccccccc}
        \hline\hline
        Nucl & $(n,l)$ & $\varepsilon_{l+}$ & $\varepsilon_{l-}$ & $\Gamma_{l+}$ & $\Gamma_{l-}$ & $\Delta B^\text{(em)}_{l+1\to l}$ & $\Delta \varepsilon^V_{l+1\to l}$ & $\Delta \varepsilon^S_{l+1\to l}$ \\
        \hline
        $^{48}\mathrm{Ca}$ & $(5,4)$ & $13.35$ & $14.39$ & $68.03$ & $71.22$ & $238.8$ & $-1.034$ & $-0.132$ \\
        $^{50}\mathrm{Ti}$ &  & $38.04$ & $41.01$ & $171.6$ & $179.9$ & $350.2$ & $-2.959$ & $-0.600$ \\
        $^{52}\mathrm{Cr}$ &  & $95.31$ & $102.83$ & $386.1$ & $405.4$ & $496.8$ & $-7.497$ & $-2.017$ \\
        $^{54}\mathrm{Fe}$ &  & $220.5$ & $238.0$ & $803.4$ & $844.5$ & $685.3$ & $-17.489$ & $-5.777$ \\
        \hline $^{86}\mathrm{Kr}$ & $(6,5)$ & $119.4$ & $130.6$ & $494.5$ & $525.3$ & $877.1$ & $-11.200$ & $-4.025$ \\
        $^{88}\mathrm{Sr}$ &  & $220.1$ & $241.1$ & $844.1$ & $897.7$ & $1089.9$ & $-20.871$ & $-8.286$ \\
        $^{90}\mathrm{Zr}$ &  & $391.8$ & $429.5$ & $1391.4$ & $1480.9$ & $1339.3$ & $-37.439$ & $-16.188$ \\
        $^{92}\mathrm{Mo}$ &  & $681.2$ & $747.0$ & $2234.6$ & $2379.0$ & $1629.2$ & $-65.267$ & $-30.438$ \\
        \hline $^{58}\mathrm{Fe}$ & $(5,4)$ & $275.6$ & $296.4$ & $931.9$ & $977.6$ & $686.1$ & $-20.666$ & $-7.254$ \\
        $^{59}\mathrm{Co}$ &  & $391.6$ & $421.5$ & $1272.4$ & $1335.6$ & $798.4$ & $-29.732$ & $-11.179$ \\
        $^{60}\mathrm{Ni}$ &  & $556.5$ & $599.3$ & $1729.0$ & $1815.6$ & $924.0$ & $-42.557$ & $-17.073$ \\
        \hline $^{114}\mathrm{Cd}$ & $(7,6)$ & $58.29$ & $65.10$ & $294.4$ & $317.1$ & $1131.8$ & $-6.772$ & $-2.504$ \\
        $^{115}\mathrm{In}$ &  & $74.86$ & $83.67$ & $367.1$ & $395.6$ & $1229.5$ & $-8.754$ & $-3.356$ \\
        $^{116}\mathrm{Sn}$ &  & $96.55$ & $107.96$ & $458.1$ & $494.1$ & $1333.6$ & $-11.337$ & $-4.502$ \\
        \hline $^{208}\mathrm{Pb}$ & $(9,8)$ & $71.48$ & $82.66$ & $388.5$ & $429.5$ & $2232.1$ & $-11.060$ & $-5.096$ \\
        $^{209}\mathrm{Bi}$ &  & $89.09$ & $102.95$ & $466.3$ & $515.5$ & $2343.7$ & $-13.706$ & $-6.451$ \\
        \hline\hline
    \end{tabular*}
    \label{tab:fullresultwycech}
\end{table}

\section{Comparison with the Perturbative Analysis}
In the main body we discuss the spin-orbit splittings and their correlations using the fully integrated Dirac equation.
In this Appendix, let us examine whether this behaviour can be found in the case when we use a somewhat simpler model.

We begin with the Coulomb potential in the point-charge approximation, assuming that the system obeys a non-relativistic, spin-independent Schr\"{o}dinger equation.  
Under these assumptions the radial wave-function $R_{nl}(r)$ is obtained as an analytic solution and may be written in a familiar form
\begin{equation}
    R_{nl}(r)=\mathcal{N}_{nl} \qty(\frac{2Zr}{a_0}n)^l e^{-Zr/a_0n} L^{2l+1}_{n-l-1} \qty(\frac{2Zr}{a_0n})
\end{equation}
where $\mathcal{N}_{nl}$ is the normalization factor
\begin{equation}
    \mathcal{N}_{nl} = \frac{2}{n^2}\sqrt{\frac{Z^3(n-l-1)!}{a_0^3(n+l)!}},
\end{equation}
and $a_0$ is the Bohr length $a_0 = \hbar c/(\alpha m_{\bar{p}}c^2)\approx 28.8\,\fm$.
Considering the perturbation of the optical potential to the first order, the level width can be derived as the overlaps between the wave function and the imaginary part of the optical potential;
\begin{equation}
    \Gamma = \int \dd r\,r^2\Im[V_\text{opt}(r)]\,|R_{nl}(r)|^{2}.
\end{equation}
The spin-orbit correction to the single-particle energy can be calculated in a similar way;
\begin{equation}
    \Delta E_{ls} = -\int\dd r\,\frac{r}{4m^2}\dv{}{r}\Re[S(r)-V(r)]c_{LS}\abs{R_{nl}(r)}^2,
\end{equation}
to the leading order of the non-relativistic approximation of the Dirac equation for the large part of the wave function (see Eq.~\eqref{eq:nonrelaDirac}.
The coefficient $c_{LS}$ is the spin-orbit component integrated out the angle components, which can be calculated as
\begin{equation}
    c_{LS} = \frac{\hbar^2}{2}\qty[j(j+1)-l(l+1)-s(s+1)]=
    \begin{cases}
        \frac{l}{2}\hbar^2 & s = +1/2\\
        -\frac{l+1}{2}\hbar^2 & s = -1/2.
    \end{cases}
\end{equation}


\begin{figure}[tbp]
    \centering
    \includegraphics[width=0.5\columnwidth]{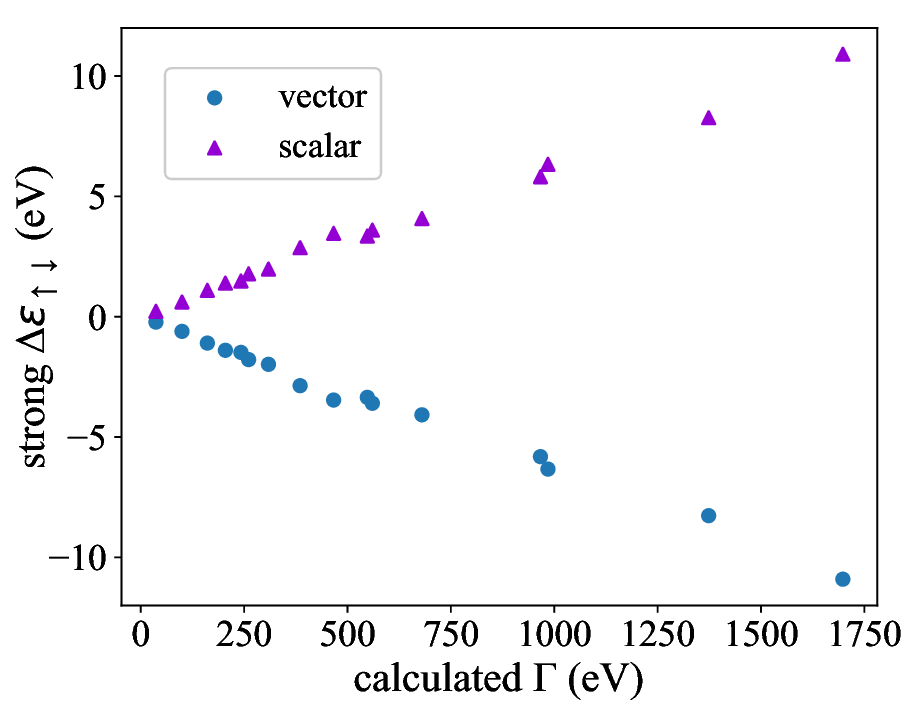}
    \caption{Calculated strong spin-orbit splittings as a function of corresponding level widths, based on the simple non-relativistic model.
    The results when the optical potential enters as vector and scalar potentials are indicated by the blue circles and violet triangles, respectively.}
    \label{fig:analytic_solution}
\end{figure}

Employing this term we can compute the strong spin-orbit splitting.
In Figure \ref{fig:analytic_solution}, calculated spin-orbit splittings as functions of level widths are plotted for the nuclei examined in the body.
For the optical potential Friedman's best fit $b_0=1.3+1.9i$ is applied, while the selected quantum numbers $(n,l)$ are same with Table \ref{tab:fullresultfriedman}.
The figure reveals that a pronounced linear correlation can be found between the strong spin-orbit splittings and level shifts or widths, even at the analytic level.  
The magnitude of the strong spin-orbit splittings differs from the numerical results, and for the scalar potential case, even the sign is opposite (see Table \ref{tab:fullresultfriedman}). 
However, when focusing on the difference between the results obtained with the scalar and vector potential cases, one finds that the values are quite close to those shown in Figure \ref{fig:shift_width_split_swave}.  
This suggests the presence of contributions that shift both results in the same direction.
For instance, as a relativistic effect, the contribution from the small component $F(r)$ may be considered, but since this accounts for only about $5\%$ of the total wave function, it is unlikely to produce a significant difference.  
On the other hand, because the Friedman potential possesses a large imaginary part, $b_0 = 1.3 + 1.9i$, the magnitude of the splittings may change through higher-order perturbative effects.

In summary, even for the strong spin-orbit splittings calculated within perturbation theory, one can observe the linear dependence on the level width, as confirmed in the main text.  
However, to evaluate their magnitude, it is necessary to go beyond a simple perturbative model and solve the Dirac equation accurately.

\bibliographystyle{ptephy}
\bibliography{antiproton}

@article{aumann2022,
  title = {{{PUMA}}, {{antiProton}} Unstable Matter Annihilation},
  author = {Aumann, T. and Bartmann, W. and {Boine-Frankenheim}, O. and Bouvard, A. and Broche, A. and Butin, F. and Calvet, D. and Carbonell, J. and Chiggiato, P. and De Gersem, H. and De Oliveira, R. and Dobers, T. and Ehm, F. and Somoza, J. Ferreira and Fischer, J. and Fraser, M. and Friedrich, E. and Frotscher, A. and {Gomez-Ramos}, M. and Grenard, J.-L. and Hobl, A. and Hupin, G. and Husson, A. and Indelicato, P. and Johnston, K. and Klink, C. and Kubota, Y. and Lazauskas, R. and {Malbrunot-Ettenauer}, S. and Marsic, N. and O M{\"u}ller, W. F. and Naimi, S. and Nakatsuka, N. and Necca, R. and Neidherr, D. and Neyens, G. and Obertelli, A. and Ono, Y. and Pasinelli, S. and Paul, N. and Pollacco, E. C. and Rossi, D. and Scheit, H. and Schlaich, M. and Schmidt, A. and Schweikhard, L. and Seki, R. and Sels, S. and Siesling, E. and Uesaka, T. and Vil{\'e}n, M. and Wada, M. and Wienholtz, F. and Wycech, S. and Zacarias, S.},
  year = {2022},
  journal = {Eur. Phys. J. A},
  volume = {58},
  pages = {88},
  issn = {1434-601X},
  url = {https://doi.org/10.1140/epja/s10050-022-00713-x}
}

@article{batty1981,
  title = {Optical-Model Analysis of Exotic Atom Data: ({{II}}). {{Antiprotonic}} and Sigma Atoms},
  shorttitle = {Optical-Model Analysis of Exotic Atom Data},
  author = {Batty, C. J.},
  year = {1981},
  journal = {Nuclear Physics A},
  volume = {372},
  pages = {433--444},
  issn = {0375-9474},
  url = {https://www.sciencedirect.com/science/article/pii/0375947481900452}
}

@article{batty1987,
  title = {Optical Model Analysis of Antiprotonic Oxygen Atom Data},
  author = {Batty, C. J.},
  year = {1987},
  journal = {Physics Letters B},
  volume = {189},
  pages = {393--396},
  issn = {0370-2693},
  url = {https://www.sciencedirect.com/science/article/pii/0370269387906472}
}

@article{batty1989,
  title = {Antiprotonic-Hydrogen Atoms},
  author = {Batty, C. J.},
  year = {1989},
  journal = {Rep. Prog. Phys.},
  volume = {52},
  pages = {1165},
  issn = {0034-4885},
  url = {https://dx.doi.org/10.1088/0034-4885/52/10/001}
}

@incollection{batty1989a,
  title = {Experimental {{Methods}} for {{Studying Nuclear Density Distributions}}},
  booktitle = {Advances in {{Nuclear Physics}}},
  author = {Batty, C. J. and Friedman, E. and Gils, H. J. and Rebel, H.},
  editor = {Negele, J. W. and Vogt, Erich},
  year = {1989},
  pages = {1--188},
  publisher = {Springer US},
  address = {Boston, MA},
  url = {https://doi.org/10.1007/978-1-4613-9907-0_1},
  isbn = {978-1-4613-9907-0}
}

@article{batty1995b,
  title = {Density-Dependent P{$<$}math{$><$}mtext{$>$}p{$<$}/Mtext{$><$}/Math{$>$}-Nucleus Optical Potentials from Global Fits to P{$<$}math{$><$}mtext{$>$}p{$<$}/Mtext{$><$}/Math{$>$} Atom Data},
  author = {Batty, C. J. and Friedman, E. and Gal, A.},
  year = {1995},
  journal = {Nuclear Physics A},
  volume = {592},
  pages = {487--512},
  issn = {0375-9474},
  url = {https://www.sciencedirect.com/science/article/pii/037594749500308N}
}

@article{batty1997a,
  title = {Strong Interaction Physics from Hadronic Atoms},
  author = {Batty, C. J. and Friedman, E. and Gal, A.},
  year = {1997},
  journal = {Physics Reports},
  volume = {287},
  pages = {385--445},
  issn = {0370-1573},
  url = {https://www.sciencedirect.com/science/article/pii/S0370157397000112}
}

@article{borie1983,
  title = {Vacuum Polarization Corrections and Spin-Orbit Splitting in Antiprotonic Atoms},
  author = {Borie, E.},
  year = {1983},
  journal = {Phys. Rev. A},
  volume = {28},
  pages = {555--558},
  publisher = {American Physical Society},
  url = {https://link.aps.org/doi/10.1103/PhysRevA.28.555}
}

@article{cote1982,
  title = {Nucleon-{{Antinucleon Optical Potential}}},
  author = {C{\^o}t{\'e}, J. and Lacombe, M. and Loiseau, B. and Moussallam, B. and Mau, R. Vinh},
  year = {1982},
  journal = {Phys. Rev. Lett.},
  volume = {48},
  pages = {1319--1322},
  publisher = {American Physical Society},
  url = {https://link.aps.org/doi/10.1103/PhysRevLett.48.1319}
}

@article{devries1987,
  title = {Nuclear Charge-Density-Distribution Parameters from Elastic Electron Scattering},
  author = {De Vries, H. and De Jager, C. W. and De Vries, C.},
  year = {1987},
  journal = {Atomic Data and Nuclear Data Tables},
  volume = {36},
  pages = {495--536},
  issn = {0092-640X},
  url = {https://www.sciencedirect.com/science/article/pii/0092640X87900131}
}

@article{doser2022,
  title = {Antiprotonic Bound Systems},
  author = {Doser, M.},
  year = {2022},
  journal = {Progress in Particle and Nuclear Physics},
  volume = {125},
  pages = {103964},
  issn = {0146-6410},
  url = {https://www.sciencedirect.com/science/article/pii/S0146641022000254}
}

@article{ficek2018,
  title = {Constraints on {{Exotic Spin-Dependent Interactions Between Matter}} and {{Antimatter}} from {{Antiprotonic Helium Spectroscopy}}},
  author = {Ficek, Filip and Fadeev, Pavel and Flambaum, Victor V. and Jackson Kimball, Derek F. and Kozlov, Mikhail G. and Stadnik, Yevgeny V. and Budker, Dmitry},
  year = {2018},
  journal = {Phys. Rev. Lett.},
  volume = {120},
  pages = {183002},
  publisher = {American Physical Society},
  url = {https://link.aps.org/doi/10.1103/PhysRevLett.120.183002}
}

@article{Friedman2008,
    author = "Friedman, E.",
    title = "{Unified approach to nuclear densities from exotic atoms}",
    archivePrefix = "arXiv",
    primaryClass = "nucl-th",
    doi = "10.1007/s10751-009-0066-x",
    journal = "Hyperfine Interact.",
    volume = "193",
    number = "1-3",
    pages = "33--38",
    year = "2009"
}

@article{friedman2004,
  title = {Antiprotonic Potentials from Global Fits to the {{PS209}} Data},
  author = {Friedman, E and Gal, A},
  year = {2004},
  journal = {Nuclear Instruments and Methods in Physics Research Section B: Beam Interactions with Materials and Atoms},
  series = {Low {{Energy Antiproton Physics}} ({{LEAP}}'03)},
  volume = {214},
  pages = {160--163},
  issn = {0168-583X},
  url = {https://www.sciencedirect.com/science/article/pii/S0168583X03017634}
}

@article{friedman2005,
  title = {Antiproton--Nucleus Potentials from Global Fits to Antiprotonic {{X-rays}} and Radiochemical Data},
  author = {Friedman, E. and Gal, A. and Mare{\v s}, J.},
  year = {2005},
  journal = {Nuclear Physics A},
  volume = {761},
  pages = {283--295},
  issn = {0375-9474},
  url = {https://www.sciencedirect.com/science/article/pii/S0375947405010146}
}

@article{friedman2007a,
  title = {In-Medium Nuclear Interactions of Low-Energy Hadrons},
  author = {Friedman, E. and Gal, A.},
  year = {2007},
  journal = {Physics Reports},
  volume = {452},
  pages = {89--153},
  issn = {0370-1573},
  url = {https://www.sciencedirect.com/science/article/pii/S0370157307003353}
}

@article{friedman2013,
  title = {Kaonic Atoms and In-Medium {{K}}-{{N}}{$<$}math{$><$}msup Is="true"{$><$}mrow Is="true"{$><$}mi Is="true"{$>$}{{K}}{$<$}/Mi{$><$}/Mrow{$><$}mrow Is="true"{$><$}mo Is="true"{$>-<$}/Mo{$><$}/Mrow{$><$}/Msup{$><$}mi Is="true"{$>$}{{N}}{$<$}/Mi{$><$}/Math{$>$} Amplitudes {{II}}: {{Interplay}} between Theory and Phenomenology},
  shorttitle = {Kaonic Atoms and In-Medium {{K}}-{{N}}{$<$}math{$><$}msup Is="true"{$><$}mrow Is="true"{$><$}mi Is="true"{$>$}{{K}}{$<$}/Mi{$><$}/Mrow{$><$}mrow Is="true"{$><$}mo Is="true"{$>-<$}/Mo{$><$}/Mrow{$><$}/Msup{$><$}mi Is="true"{$>$}{{N}}{$<$}/Mi{$><$}/Math{$>$} Amplitudes {{II}}},
  author = {Friedman, E. and Gal, A.},
  year = {2013},
  journal = {Nuclear Physics A},
  volume = {899},
  pages = {60--75},
  issn = {0375-9474},
  url = {https://www.sciencedirect.com/science/article/pii/S0375947413000262}
}

@article{friedman2014,
  title = {Testing In-Medium {{{\emph{$\pi$N}}}} Dynamics on Pionic Atoms},
  author = {Friedman, E. and Gal, A.},
  year = {2014},
  journal = {Nuclear Physics A},
  series = {Special {{Issue Dedicated}} to the {{Memory}} of {{Gerald E Brown}} (1926-2013)},
  volume = {928},
  pages = {128--137},
  issn = {0375-9474},
  url = {https://www.sciencedirect.com/science/article/pii/S0375947414001511}
}

@article{friedman2015,
  title = {Antinucleon--Nucleus Interaction near Threshold from the {{Paris N}}{\textasciimacron}{{N}}{$<$}math{$><$}mover Accent="true" Is="true"{$><$}mrow Is="true"{$><$}mi Is="true"{$>$}{{N}}{$<$}/Mi{$><$}/Mrow{$><$}mrow Is="true"{$><$}mo Stretchy="false" Is="true"{$>$}{\textasciimacron}{$<$}/Mo{$><$}/Mrow{$><$}/Mover{$><$}mi Is="true"{$>$}{{N}}{$<$}/Mi{$><$}/Math{$>$} Potential},
  author = {Friedman, E. and Gal, A. and Loiseau, B. and Wycech, S.},
  year = {2015},
  journal = {Nuclear Physics A},
  volume = {943},
  pages = {101--116},
  issn = {0375-9474},
  url = {https://www.sciencedirect.com/science/article/pii/S0375947415001992}
}

@article{friedman2019,
  title = {The Pion-Nucleon {\emph{{$\sigma$}}} Term from Pionic Atoms},
  author = {Friedman, E. and Gal, A.},
  year = {2019},
  journal = {Physics Letters B},
  volume = {792},
  pages = {340--344},
  issn = {0370-2693},
  url = {https://www.sciencedirect.com/science/article/pii/S0370269319301935}
}

@article{friedman2019a,
  title = {Overview of Strong Interaction from Kaonic Atoms},
  author = {Friedman, E.},
  year = {2019},
  journal = {EPJ Web Conf.},
  volume = {199},
  pages = {01013},
  publisher = {EDP Sciences},
  issn = {2100-014X},
  url = {https://www.epj-conferences.org/articles/epjconf/abs/2019/04/epjconf_meson2019_01013/epjconf_meson2019_01013.html},
  copyright = {{\copyright} The Authors, published by EDP Sciences, 2019}
}

@article{fullerton1976,
  title = {Accurate and Efficient Methods for the Evaluation of Vacuum-Polarization Potentials of Order \${{Z}}{\textbackslash}ensuremath\{{\textbackslash}alpha\}\$ and \${{Z}}\{{\textbackslash}ensuremath\{{\textbackslash}alpha\}\}{\textasciicircum}\{2\}\$},
  author = {Fullerton, L. Wayne and Rinker, G. A.},
  year = {1976},
  journal = {Phys. Rev. A},
  volume = {13},
  pages = {1283--1287},
  publisher = {American Physical Society},
  url = {https://link.aps.org/doi/10.1103/PhysRevA.13.1283}
}

@article{gaitanos2011,
  title = {How Deep Is the Antinucleon Optical Potential at {{FAIR}} Energies},
  author = {Gaitanos, T. and Kaskulov, M. and Lenske, H.},
  year = {2011},
  journal = {Physics Letters B},
  volume = {703},
  pages = {193--198},
  issn = {0370-2693},
  url = {https://www.sciencedirect.com/science/article/pii/S0370269311008896}
}

@article{gotta1999,
  title = {Balmer {\emph{{$\alpha$}}} Transitions in Antiprotonic Hydrogen and Deuterium},
  author = {Gotta, D. and Anagnostopoulos, D. F. and Augsburger, M. and Borchert, G. and Castelli, C. and Chatellard, D. and Egger, J. -P. and {El-Khoury}, P. and Gorke, H. and Hauser, P. and Indelicato, P. and Kirch, K. and Lenz, S. and Nelms, N. and Rashid, K. and Siems, {\relax Th}. and Simons, L. M.},
  year = {1999},
  journal = {Nuclear Physics A},
  volume = {660},
  pages = {283--321},
  issn = {0375-9474},
  url = {https://www.sciencedirect.com/science/article/pii/S0375947499003851}
}

@article{gotta2004,
  title = {Precision Spectroscopy of Light Exotic Atoms},
  author = {Gotta, D.},
  year = {2004},
  journal = {Progress in Particle and Nuclear Physics},
  volume = {52},
  pages = {133--195},
  issn = {0146-6410},
  url = {https://www.sciencedirect.com/science/article/pii/S0146641003001121}
}

@article{hartmann2001a,
  title = {Nucleon Density in the Nuclear Periphery Determined with Antiprotonic x Rays: {{Calcium}} Isotopes},
  shorttitle = {Nucleon Density in the Nuclear Periphery Determined with Antiprotonic x Rays},
  author = {Hartmann, F. J. and Schmidt, R. and Ketzer, B. and {von Egidy}, T. and Wycech, S. and Smola{\'n}czuk, R. and Czosnyka, T. and Jastrz{\c e}bski, J. and Kisieli{\'n}ski, M. and Lubi{\'n}ski, P. and Napiorkowski, P. and Pie{\'n}kowski, L. and Trzci{\'n}ska, A. and K{\l}os, B. and Gulda, K. and Kurcewicz, W. and Widmann, E.},
  year = {2001},
  journal = {Phys. Rev. C},
  volume = {65},
  pages = {014306},
  publisher = {American Physical Society},
  url = {https://link.aps.org/doi/10.1103/PhysRevC.65.014306}
}

@misc{higuchi2025,
  title = {Precision {{Spectroscopy}} of {{Antiprotonic Atoms}} for {{Investigation}} of {{Low-energy Antinucleon-nucleus Interactions}}},
  author = {Higuchi, Takashi and Fujioka, Hiroyuki},
  year = {2025},
  eprint = {2501.08759},
  primaryclass = {nucl-ex},
  publisher = {arXiv},
  url = {http://arxiv.org/abs/2501.08759},
  archiveprefix = {arXiv}
}

@article{jastrzebski2004,
  title = {Neutron Density Distributions from Antiprotonic Atoms Compared with Hadron Scattering Data},
  author = {Jastrz{\c e}bski, J. and Trzci{\'n}ska, A. and Lubi{\'n}ski, P. and K{\l}os, B. and Hartmann, F. J. and {von EGIDY}, T. and Wycech, S.},
  year = {2004},
  journal = {Int. J. Mod. Phys. E},
  volume = {13},
  pages = {343--351},
  publisher = {World Scientific Publishing Co.},
  issn = {0218-3013},
  url = {https://www.worldscientific.com/doi/abs/10.1142/S0218301304002168}
}

@article{klempt2005,
  title = {The Antinucleon--Nucleon Interaction at Low Energy: {{Annihilation}} Dynamics},
  shorttitle = {The Antinucleon--Nucleon Interaction at Low Energy},
  author = {Klempt, Eberhard and Batty, Chris and Richard, Jean-Marc},
  year = {2005},
  journal = {Physics Reports},
  volume = {413},
  pages = {197--317},
  issn = {0370-1573},
  url = {https://www.sciencedirect.com/science/article/pii/S0370157305001055}
}

@article{klos2004,
  title = {Strong Interaction and {{E}} 2 Effect in Even- {{A}} Antiprotonic {{Te}} Atoms},
  author = {K{\l}os, B. and Wycech, S. and Trzci{\'n}ska, A. and Jastrz{\c e}bski, J. and Czosnyka, T. and Kisieli{\'n}ski, M. and Lubi{\'n}ski, P. and Napiorkowski, P. and Pie{\'n}kowski, L. and Hartmann, F. J. and Ketzer, B. and Schmidt, R. and Von Egidy, T. and Cugnon, J. and Gulda, K. and Kurcewicz, W. and Widmann, E.},
  year = {2004},
  journal = {Phys. Rev. C},
  volume = {69},
  pages = {044311},
  issn = {0556-2813, 1089-490X},
  url = {https://link.aps.org/doi/10.1103/PhysRevC.69.044311},
  copyright = {http://link.aps.org/licenses/aps-default-license}
}

@article{klos2007,
  title = {Neutron Density Distributions from Antiprotonic {{Pb}} 208 and {{Bi}} 209 Atoms},
  author = {K{\l}os, B. and Trzci{\'n}ska, A. and Jastrz{\k e}bski, J. and Czosnyka, T. and Kisieli{\'n}ski, M. and Lubi{\'n}ski, P. and Napiorkowski, P. and Pie{\'n}kowski, L. and Hartmann, F. J. and Ketzer, B. and Ring, P. and Schmidt, R. and Egidy, T. Von and Smola{\'n}czuk, R. and Wycech, S. and Gulda, K. and Kurcewicz, W. and Widmann, E. and Brown, B. A.},
  year = {2007},
  journal = {Phys. Rev. C},
  volume = {76},
  pages = {014311},
  issn = {0556-2813, 1089-490X},
  url = {https://link.aps.org/doi/10.1103/PhysRevC.76.014311},
  copyright = {http://link.aps.org/licenses/aps-default-license}
}

@article{kreissl1988a,
  title = {First Direct Observation of Strong Interaction Spin-Orbit Effects in Antiprotonic Atoms},
  author = {Kreissl, A. and Hancock, A. D. and Koch, H. and K{\"o}hler, {\relax Th}. and Poth, H. and Raich, U. and Rohmann, D. and Wolf, A. and Tauscher, L. and Nilsson, A. and Suffert, M. and Chardalas, M. and Dedoussis, S. and Daniel, H. and {von Egidy}, T. and Hartmann, F. J. and Kanert, W. and Plendl, H. and Schmidt, G. and Reidy, J. J.},
  year = {1988},
  journal = {Z. Physik A - Atomic Nuclei},
  volume = {329},
  pages = {235--241},
  issn = {0939-7922},
  url = {https://doi.org/10.1007/BF01283780}
}

@article{lubinski1998,
  title = {Composition of the Nuclear Periphery from Antiproton Absorption},
  author = {Lubi{\'n}ski, P. and Jastrz{\c e}bski, J. and Trzci{\'n}ska, A. and Kurcewicz, W. and Hartmann, F. J. and Schmid, W. and {von Egidy}, T. and Smola{\'n}czuk, R. and Wycech, S.},
  year = {1998},
  journal = {Phys. Rev. C},
  volume = {57},
  pages = {2962--2973},
  publisher = {American Physical Society},
  url = {https://link.aps.org/doi/10.1103/PhysRevC.57.2962}
}

@article{nishi2023,
  title = {Chiral Symmetry Restoration at High Matter Density Observed in Pionic Atoms},
  author = {Nishi, Takahiro and Itahashi, Kenta and Ahn, DeukSoon and Berg, Georg P. A. and Dozono, Masanori and Etoh, Daijiro and Fujioka, Hiroyuki and Fukuda, Naoki and Fukunishi, Nobuhisa and Geissel, Hans and Haettner, Emma and Hashimoto, Tadashi and Hayano, Ryugo S. and Hirenzaki, Satoru and Horii, Hiroshi and Ikeno, Natsumi and Inabe, Naoto and Iwasaki, Masahiko and Kameda, Daisuke and Kisamori, Keichi and Kiyokawa, Yu and Kubo, Toshiyuki and Kusaka, Kensuke and Matsushita, Masafumi and Michimasa, Shin'ichiro and Mishima, Go and Miya, Hiroyuki and Murai, Daichi and Nagahiro, Hideko and Niikura, Megumi and {Nose-Togawa}, Naoko and Ota, Shinsuke and Sakamoto, Naruhiko and Sekiguchi, Kimiko and Shiokawa, Yuta and Suzuki, Hiroshi and Suzuki, Ken and Takaki, Motonobu and Takeda, Hiroyuki and Tanaka, Yoshiki K. and Uesaka, Tomohiro and Wada, Yasumori and Watanabe, Atomu and Watanabe, Yuni N. and Weick, Helmut and Yamakami, Hiroki and Yanagisawa, Yoshiyuki and Yoshida, Koichi},
  year = {2023},
  journal = {Nat. Phys.},
  volume = {19},
  pages = {788--793},
  publisher = {Nature Publishing Group},
  issn = {1745-2481},
  url = {https://www.nature.com/articles/s41567-023-02001-x},
  copyright = {2023 The Author(s), under exclusive licence to Springer Nature Limited}
}

@article{schmidt2003,
  title = {Nucleon Density in the Nuclear Periphery Determined with Antiprotonic x Rays: {{Cadmium}} and Tin Isotopes},
  shorttitle = {Nucleon Density in the Nuclear Periphery Determined with Antiprotonic x Rays},
  author = {Schmidt, R. and Trzci{\'n}ska, A. and Czosnyka, T. and {von Egidy}, T. and Gulda, K. and Hartmann, F. J. and Jastrz{\c e}bski, J. and Ketzer, B. and Kisieli{\'n}ski, M. and K{\l}os, B. and Kurcewicz, W. and Lubi{\'n}ski, P. and Napiorkowski, P. and Pie{\'n}kowski, L. and Smola{\'n}czuk, R. and Widmann, E. and Wycech, S.},
  year = {2003},
  journal = {Phys. Rev. C},
  volume = {67},
  pages = {044308},
  publisher = {American Physical Society},
  url = {https://link.aps.org/doi/10.1103/PhysRevC.67.044308}
}

@article{smorra2017,
  title = {A Parts-per-Billion Measurement of the Antiproton Magnetic Moment},
  author = {Smorra, C. and Sellner, S. and Borchert, M. J. and Harrington, J. A. and Higuchi, T. and Nagahama, H. and Tanaka, T. and Mooser, A. and Schneider, G. and Bohman, M. and Blaum, K. and Matsuda, Y. and Ospelkaus, C. and Quint, W. and Walz, J. and Yamazaki, Y. and Ulmer, S.},
  year = {2017},
  journal = {Nature},
  volume = {550},
  pages = {371--374},
  publisher = {Nature Publishing Group},
  issn = {1476-4687},
  url = {https://www.nature.com/articles/nature24048},
  copyright = {2017 The Author(s)}
}

@article{trzcinska2001a,
  title = {Information on Antiprotonic Atoms and the Nuclear Periphery from the {{PS209}} Experiment},
  author = {Trzci{\'n}ska, A. and Jastrz{\c e}bski, J. and Czosnyka, T. and {von Egidy}, T. and Gulda, K. and Hartmann, F. J. and Iwanicki, J. and Ketzer, B. and Kisieli{\'n}ski, M. and K{\l}os, B. and Kurcewicz, W. and Lubi{\'n}ski, P. and Napiorkowski, P. J. and Pie{\'n}kowski, L. and Schmidt, R. and Widmann, E.},
  year = {2001},
  journal = {Nuclear Physics A},
  series = {Sixth {{Biennial Conference}} on {{Low-Energy Antiproton Physics}}},
  volume = {692},
  pages = {176--181},
  issn = {0375-9474},
  url = {https://www.sciencedirect.com/science/article/pii/S0375947401011769}
}

@article{trzcinska2009,
  title = {Nuclear Periphery Studied with Antiprotonic Atoms},
  author = {Trzci{\'n}ska, A. and {for PS209 collaboration}},
  year = {2009},
  journal = {Hyperfine Interact},
  volume = {194},
  pages = {271--276},
  issn = {1572-9540},
  url = {https://doi.org/10.1007/s10751-009-0078-6}
}

@article{wycech1993,
  title = {Antiprotonic Atoms with Heavy Nuclei},
  author = {Wycech, S. and Hartmann, F. J. and Daniel, H. and Kanert, W. and Plendl, H. S. and {von Egidy}, T. and Reidy, J. J. and Nicholas, M. and Redmond, L. A. and Koch, H. and Kreissl, A. and Poth, H. and Rohmann, D.},
  year = {1993},
  journal = {Nuclear Physics A},
  volume = {561},
  pages = {607--627},
  issn = {0375-9474},
  url = {https://www.sciencedirect.com/science/article/pii/0375947493900689}
}

@article{wycech1996c,
  title = {Antiprotonic Studies of Nuclear Neutron Halos},
  author = {Wycech, S. and Skalski, J. and Smola{\'n}czuk, R. and Dobaczewski, J. and Rook, J. R.},
  year = {1996},
  journal = {Phys. Rev. C},
  volume = {54},
  pages = {1832--1842},
  issn = {0556-2813, 1089-490X},
  url = {https://link.aps.org/doi/10.1103/PhysRevC.54.1832},
  copyright = {http://link.aps.org/licenses/aps-default-license}
}

@article{wycech2007,
  title = {Nuclear Surface Studies with Antiprotonic Atom x Rays},
  author = {Wycech, S. and Hartmann, F. J. and Jastrz{\k e}bski, J. and K{\l}os, B. and Trzci{\'n}ska, A. and von Egidy, T.},
  year = {2007},
  journal = {Phys. Rev. C},
  volume = {76},
  pages = {034316},
  publisher = {American Physical Society},
  url = {https://link.aps.org/doi/10.1103/PhysRevC.76.034316}
}

@misc{yoshimura2024a,
  title = {Interrelation between \${\textbackslash}bar\{p\}\$-{{Ca Atom Spectra}} and {{Nuclear Density Profiles}}},
  author = {Yoshimura, Kenta and Yasunaga, Shunsuke and Jido, Daisuke and {Yamagata-Sekihara}, Junko and Hirenzaki, Satoru},
  year = {2024},
  eprint = {2408.14760},
  primaryclass = {nucl-th},
  publisher = {arXiv},
  url = {http://arxiv.org/abs/2408.14760},
  archiveprefix = {arXiv}
}

@article{zotero-626,
  journal = {{Tables} of {Nuclear Data}},
  pages = {https://wwwndc.jaea.go.jp/NuC/}
}

@article{fricke1995,
  title = {Nuclear {{Ground State Charge Radii}} from {{Electromagnetic Interactions}}},
  author = {Fricke, G. and Bernhardt, C. and Heilig, K. and Schaller, L. A. and Schellenberg, L. and Shera, E. B. and Dejager, C. W.},
  year = {1995},
  journal = {Atomic Data and Nuclear Data Tables},
  volume = {60},
  pages = {177--285},
  issn = {0092-640X},
  url = {https://www.sciencedirect.com/science/article/pii/S0092640X85710078}
}

@misc{gustafsson2025,
  title = {Stability Conditions for Bound States in Antiprotonic Atoms},
  author = {Gustafsson, Fredrik Parnefjord and P{\k e}cak, Daniel and Sowi{\'n}ski, Tomasz},
  year = {2025},
  eprint = {2502.17192},
  primaryclass = {physics},
  publisher = {arXiv},
  url = {http://arxiv.org/abs/2502.17192},
  archiveprefix = {arXiv}
}

@article{friedman2014a,
  title = {Antineutron and Antiproton Nuclear Interactions at Very Low Energies},
  author = {Friedman, E.},
  year = {2014},
  journal = {Nuclear Physics A},
  volume = {925},
  pages = {141--149},
  issn = {0375-9474},
  url = {https://www.sciencedirect.com/science/article/pii/S0375947414000463}
}

@article{COHEN1995221,
title = {QCD sum rules and applications to nuclear physics},
journal = {Progress in Particle and Nuclear Physics},
volume = {35},
pages = {221-298},
year = {1995},
issn = {0146-6410},
doi = {https://doi.org/10.1016/0146-6410(95)00043-I},
url = {https://www.sciencedirect.com/science/article/pii/014664109500043I},
author = {T.D. Cohen and R.J. Furnstahl and D.K. Griegel and Xuemin Jin}
}

@article{HRTANKOVA2016197,
title = {Interaction of antiprotons with nuclei},
journal = {Nuclear Physics A},
volume = {945},
pages = {197-215},
year = {2016},
issn = {0375-9474},
doi = {https://doi.org/10.1016/j.nuclphysa.2015.10.005},
url = {https://www.sciencedirect.com/science/article/pii/S0375947415002286},
author = {Jaroslava Hrtánková and Jiří Mareš}
}

@article{poth1985,
author="Poth, H.",
editor="von Geramb, H. V.",
title="Recent results from antiprotonic atoms at LEAR",
journal="Medium Energy Nucleon and Antinucleon Scattering",
year="1985",
publisher="Springer Berlin Heidelberg",
address="Berlin, Heidelberg",
pages="357--367",
isbn="978-3-540-39739-7"
}

@article{mishustin2005,
  title = {Antibaryons bound in nuclei},
  author = {Mishustin, I. N. and Satarov, L. M. and B\"urvenich, T. J. and St\"ocker, H. and Greiner, W.},
  journal = {Phys. Rev. C},
  volume = {71},
  issue = {3},
  pages = {035201},
  numpages = {32},
  year = {2005},
  month = {Mar},
  publisher = {American Physical Society},
  doi = {10.1103/PhysRevC.71.035201},
  url = {https://link.aps.org/doi/10.1103/PhysRevC.71.035201}
}

@article{lisboa2010,
  title = {Spin and pseudospin symmetries in the antinucleon spectrum of nuclei},
  author = {Lisboa, R. and Malheiro, M. and Alberto, P. and Fiolhais, M. and de Castro, A. S.},
  journal = {Phys. Rev. C},
  volume = {81},
  issue = {6},
  pages = {064324},
  numpages = {8},
  year = {2010},
  month = {Jun},
  publisher = {American Physical Society},
  doi = {10.1103/PhysRevC.81.064324},
  url = {https://link.aps.org/doi/10.1103/PhysRevC.81.064324}
}

\end{document}